\documentclass[a4paper,10pt]{article}

\usepackage[english]{babel}
\usepackage[utf8x]{inputenc}
\usepackage[T1]{fontenc}

\usepackage[a4paper,top=3cm,bottom=2cm,left=3cm,right=3cm,marginparwidth=1.75cm]{geometry}

\usepackage{amsmath,amsfonts,amssymb}
\usepackage{graphicx}
\usepackage{amsthm}
\usepackage{xcolor}

\newtheorem{theorem}{\textsc{Theorem}}
\newtheorem{lemma}{\textsc{Lemma}}
\newtheorem{definition}{\textsc{Definition}}

\newtheorem{corollary}{\textsc{Corollary}}[section]

\makeatletter
\def\@maketitle{%
  \newpage
  \null
  \vskip 2em%
  \begin{center}%
  \let \footnote \thanks
    {\huge \@title \par}%
    \vskip 1.5em%
    {\large
      \lineskip .5em%
      \begin{tabular}[t]{c}%
        \@author
      \end{tabular}\par}%
    \vskip 1em%
  \end{center}%
  \par
  \vskip 1.5em
  \vspace{1cm}}
\makeatother

\begin{document}

\title{Classification of bounded travelling wave solutions for the Dullin-Gottwald-Holm equation\vspace{1cm}}

\author{\textbf{Priscila~Leal~da~Silva}\\
        \large Departamento de Matem\'atica\\
        Universidade Federal de S\~ao Carlos\\
        Rodovia Washington Luís km 235 - SP-310\\
        São Carlos, São Paulo, Brazil\\
        \textit{pri.leal.silva@gmail.com}
        }

\maketitle

\renewcommand{\abstractname}{\vspace{-\baselineskip}}

\begin{abstract}
\centering\begin{minipage}{\dimexpr\paperwidth-10cm}
\textbf{Abstract:} In this paper we classify all bounded travelling wave solutions for the integrable Dullin-Gottwald-Holm equation. It is shown that it decomposes in two known cases: the Camassa-Holm and the Korteweg-de Vries equation. For the former, the classification is similar to the one presented in [J. Lenells, Travelling wave solutions of the Camassa–Holm equation, \emph{J. Diff. Eq.}, \textbf{v. 217}, 393--430, (2005)], while for the latter it is only possible to obtain smooth solutions.

\vspace{0.5cm}
\textbf{Keywords:} Travelling waves, Dullin-Gottwald-Holm equation, integrable equations.
\end{minipage}
\end{abstract}

\bigskip

\section{Introduction}

In \cite{CH}, Camassa and Holm deduced the so called Camassa-Holm equation
\begin{align}\label{CH}
m_t + c_0u_x + 2u_xm + um_x=0,\quad m=u-u_{xx}, \quad c_0 \in \mathbb{R},
\end{align}
as a model of unidirectional propagation of water waves in a shallow regime. Equation \eqref{CH} possesses remarkable properties such as a bi-Hamiltonian formulation, Lax pairs \cite{FF}, an infinite hierarchy of conservation laws and peaked solitons (peakons) \cite{CH}, and since the work of Camassa and Holm, intense research has been dedicated to nonlocally evolutive equations of the type \eqref{CH}. In particular, we mention the Dullin-Gottwald-Holm equation
\begin{align}\label{DGH}
m_t + c_0u_x + 2u_xm + um_x + \gamma u_{xxx}=0, \quad m=u-\alpha^2u_{xx}, \quad c_0,\gamma,\alpha \in \mathbb{R},
\end{align}
a shallow water equation deduced by Dullin, Gottwald and Holm \cite{DGH} employing the same asymptotic analysis used by Camassa and Holm in \cite{CH}. Physically speaking, the constants $\alpha^2$ and $\gamma/c_0$ represent squares of length scales and $c_0$ is the linear wave speed for water at rest at spatial infinity, see \cite{DGH,DGH1}. Equation \eqref{DGH} is reduced to the famous Korteweg-de Vries (KdV) equation \cite{GGKM}
\begin{align}\label{KdV}
u_t + c_0u_x + 3uu_x + u_{xxx}=0
\end{align}
when $\alpha\to 0$ and to the Camassa-Holm equation \cite{CH} when $\gamma \to 0$. Equation \eqref{DGH} is also a bi-Hamiltonian system, with a Lax pair and an infinite number of conservation laws, and its peakon solutions can be described in terms of a finite dimensional dynamical system \cite{DGH}. Furthermore, \eqref{DGH} is known to be well-posed in Sobolev spaces \cite{LY,LY2,ZX} and Besov spaces \cite{YY}; several explicit wave solutions, including solitons obtained from the inverse scattering method \cite{AG}, have been deduced \cite{BK,HL,Ju,MHLL,XXL,YLCZ}, and multipeakon \cite{LY,LY1} and general periodic solutions \cite{LY2} are orbitally stable. Furthermore blow-up criteria \cite{Liu,Nov} and the wave breaking phenomena \cite{GN,HW,ZX} have been studied and the optimal distributed control of the viscous Dullin-Gottwald-Holm equation discussed \cite{STG,Sun}.

Of particular interest in this paper are travelling wave solutions and the works of Lenells \cite{Len} and Yin and Tian \cite{YT}, where they presented a qualitative analysis of the travelling wave solutions for the Camassa-Holm \eqref{CH} equation (more specifically, \eqref{DGH} with $\alpha=1$). Using the approach proposed by Lenells \cite{Len}, the authors of both papers managed to study a certain reduction of the Camassa-Holm equation to obtain a complete classification of all bounded wave solutions of it. Among the results we can find that smooth wave solutions for \eqref{CH} will be periodic or with decay, and for weak solutions one will obtain peakons, cuspons and composite wave solutions, for further details see \cite{Len, YT}.

Although the KdV equation \eqref{KdV} and its solutions have been widely investigated since the seminal work of Gardner, Green, Kruskal and Miura \cite{GGKM}, as far as we know the case $\alpha=0$ in \eqref{DGH} has not been considered in the literature when it comes to classification of bounded travelling waves.

For this reason, in this paper we will be interested in analysing the importance of the parameters $\alpha,c_0$ and $\gamma$ for the determination of existence of bounded wave solutions of \eqref{DGH}. We will show that although $\alpha\neq 0$ enlarges the number of wave solutions admitted by \eqref{DGH}, there is no substantial difference between $\alpha=1$ and $\alpha\neq 0,1$. Moreover, we will prove that the KdV equation \eqref{KdV} does not admit weak solutions due to the quadrature form of the equation and no monotonic solutions (kinks) will exist for any choice of $\alpha$.

In the next section we present the definitions required and prove the our results. In Section \ref{Sec3} we obtain the classification of bounded waves for the case $\alpha=0$ corresponding to the KdV equation and in Section \ref{Sec4} we finally deduce the conditions for existence of bounded travelling waves for the case $\alpha\neq 0$.

\section{Travelling wave solutions}\label{Sec2}

We are interested in classifying all types of bounded travelling wave solutions of \eqref{DGH}. For this purpose, let $u(t,x)=\phi(z)$ be a bounded wave solution of \eqref{DGH}, with $z=x-ct$, where $c$ denotes the wave speed. Equation \eqref{DGH} is then rewritten as
\begin{align*}
-c\phi' + \alpha^2c \phi''' + c_0\phi' + 3\phi\phi' + \gamma\phi''' = \alpha^2(2\phi'\phi'' + \phi\phi'''),
\end{align*}
and direct integration yields
\begin{align}\label{eq2}
-c\phi +\alpha^2c\phi''+c_0\phi + \frac{3}{2}\phi^2 + \gamma \phi'' =\alpha^2\left(\frac{(\phi')^2}{2}+\phi\phi''\right)+A,
\end{align}
where $A$ is an integration constant. 
Observe that if $\alpha=\gamma=0$, then equation \eqref{DGH} becomes a Burgers equation without viscosity, a case that will not be considered in this paper. For this reason, we will only consider the case $\gamma\neq0$.

The expression \eqref{eq2} is valid in the sense of distributions if we allow $\phi\in H^1_{loc}(\mathbb{R})$ and this leads to the following definition:

\begin{definition}\label{def1}
A function $\phi\in H^1_{loc}$ is said to be a weak travelling wave solution of \eqref{DGH} if it satisfies \eqref{eq2} in the sense of distributions.
\end{definition}

It is worth mentioning that the set of solutions provided by Definition \ref{def1} contains all smooth solutions and may have other solutions that may be only continuous. In this sense, from now on by weak solution we will mean a solution with differentiability issues.

After multiplicating \eqref{eq2} by $\phi'$ and integrating once again, it is obtained
\begin{align}\label{eq1}
\left(\phi'\right)^2=\frac{P(\phi)}{\alpha^2(c-\phi) + \gamma},
\end{align}
where $P(\phi) = \phi^2(c-c_0-\phi) + A\phi+B$ and $B$ is another constant obtained from the integration process.

Lenells in \cite{Len} presented a qualitative discussion on
\begin{align}\label{qual}
(\phi')^2 = F(\phi),
\end{align}
where $F$ denotes a rational function, and the behaviour of smooth solutions according to zeros of $F$. It can be summarized in the following items:

\begin{enumerate}
\item Whenever $F$ has two simple zeros $z_1$ and $z_2$ and $F(\phi)>0$ for $z_1<\phi<z_2$, there will exist a smooth periodic solution $\phi$ for \eqref{qual} with $z_1 = \min_{z\in\mathbb{R}} \phi(z)$ and $z_2 = \max_{z\in\mathbb{R}} \phi(z)$.

\item Whenever $F$ has a double zero $z_1$, a simple zero $z_2$ and $F(\phi)>0$ for $z_1<\phi<z_2$, then there exists a solution $\phi$ with $z_1 = \inf_{z\in\mathbb{R}} \phi(z)$, $z_2 = \sup_{z\in\mathbb{R}} \phi(z)$ and $\phi \downarrow z_1$ exponentially as $z\to \pm \infty$.

\item If $F$ has a simple zero at $z_1$ and $F(\phi)>0$ for $z_1<\phi$, then no bounded solution exists for $\phi>z_1$.
\end{enumerate}

The existence of poles in $F$ leads to existence of weak solutions. This is explained by the simple fact that if $F$ has a pole $a$, then $\phi'$ blows-up as $\phi\to a$ and, therefore, $\phi$ is not differentiable on any point $x_0$ such that $\phi(x_0)=a$. The following discussion can be drawn (see Lenells \cite{Len} for further and deeper details):




\begin{enumerate}
\item[4.] Peakons will occur when $\phi$ satisfies \eqref{qual} and suddenly changes direction at $\phi(x_0) = a$:
\begin{align*}
0\neq\lim\limits_{z \uparrow x_0} \phi'(z) = - \lim\limits_{z \downarrow x_0} \phi'(z) \neq \pm\infty.
\end{align*}
This means that the pole $a$ must be ``removable'' in terms of the zeros of $F$.

\item[5.] Cusps will occur at simple poles of $F$ that cannot be removed. When $a = \min_{z\in\mathbb{R}}\phi(z)$ (or taken as the maximum), we have
\begin{align*}
\lim\limits_{z \uparrow x_0} \phi'(z) = - \lim\limits_{z \downarrow x_0} \phi'(z) = \pm\infty
\end{align*}
and cuspon solutions will exist.
\end{enumerate}


In the case of \eqref{DGH} and the quadrature form \eqref{eq1}, it is clear that for $\alpha = 0$ no poles will exist and, therefore, no weak solutions will exist. Since the cases $\alpha\neq 0$ and $\alpha=0$ change dramatically, they will be considered separately.

\section{The KdV equation and the case $\alpha=0$}\label{Sec3}


As mentioned before, the case $\alpha=0$ refers to the KdV equation \eqref{KdV}. Setting $\alpha=0$ in \eqref{eq1} yields
\begin{align}\label{eq5}
(\phi')^2 = \frac{P(\phi)}{\gamma} = \frac{ B + A\phi + (c-c_0)\phi^2 - \phi^3}{\gamma}
\end{align}
and the function $F := P/\gamma$ has no poles. This will imply that equation \eqref{DGH} with $\alpha =0$ will only admit classical wave solutions. Observe that $F(\phi)$ has either one or three real zeros. If only one zero is real, then all possibilities discussed in Section \ref{Sec2} lead to the conclusion that no bounded solutions will exist. Suppose then $F$ has three zeros $M,m$ and $z_0$ so
\begin{align*}
F(\phi) = \frac{1}{\gamma}\phi^2(c-c_0 - \phi)+ A\phi + B = \frac{1}{\alpha} (M-\phi)(\phi-m)(\phi-z_0)
\end{align*}
with the compatibility condition $$z_0 = c-c_0-M-m.$$ Combining the possibilities for $z$ and the different cases for smooth solutions we then have one of the following:
\begin{enumerate}
\item Whenever $\gamma>0$,
	\begin{enumerate}
	\item if $z_0<m<\phi<M$, then $F(\phi) >0$ and there will be a smooth periodic solution $\phi$ of \eqref{eq5} with $$m=\min_{z\in\mathbb{R}}\phi(z)\quad \text{and}\quad M=\max_{z\in\mathbb{R}}\phi(z);$$

	\item if $z_0=m<\phi<M$, then $F(\phi)>0$ and there exists a smooth solution $\phi$ with $$m=\inf_{z\in\mathbb{R}}\phi(z),\quad M=\max_{z\in\mathbb{R}}\phi(z)$$ and $\phi \downarrow m$ exponentially as $|z|\to \infty$.
	\end{enumerate}
\item Whenever $\gamma<0$,
	\begin{enumerate}
	\item $m<\phi<M<z_0$, then $F(\phi) >0$ and there will be a smooth periodic solution $\phi$ of \eqref{eq5} with $$m=\min_{z\in\mathbb{R}}\phi(z)\quad \text{and}\quad M=\max_{z\in\mathbb{R}}\phi(z);$$

	\item if $m<\phi<M=z_0$, then $F(\phi)>0$ and there exists a smooth solution $\phi$ with $$m=\min_{z\in\mathbb{R}}\phi(z),\quad M=\sup_{z\in\mathbb{R}}\phi(z)$$ and $\phi \uparrow M$ exponentially as $|z|\to \infty$.
	\end{enumerate}
\end{enumerate}
With these two cases, we prove the following theorem for the KdV equation:

\begin{theorem}[\textsc{KdV equation}]\label{teokdv}
Let $\alpha=0$ and $\phi\in H^1_{loc}(\mathbb{R})$ be a travelling wave solution of $\eqref{DGH}$ wave speed $c$. Then it falls into one of the following cases, with $z_0=c-c_0-M-m$:
	\begin{enumerate}
	\item[$(i)$] (\textit{Smooth periodic}) Whenever $\gamma>0$ and $z_0<m<M$, there will be a smooth periodic solution $\phi(x-ct)$ of \eqref{DGH} with $m=\min_{z\in\mathbb{R}}\phi(z)$ and $M=\max_{z\in\mathbb{R}}\phi(z).$
	\item[$(ii)$] (\textit{Smooth with decay}) Whenever $\gamma>0$ and $z_0=m<M$, then there exists a smooth solution $\phi(x-ct)$ of \eqref{DGH} with $m=\inf_{z\in\mathbb{R}}\phi(z), M=\max_{z\in\mathbb{R}}\phi(z)$ and $\phi \downarrow m$ exponentially as $z\to \pm\infty$.
	\item[$(iii)$] (\textit{Smooth periodic:}) Whenever $\gamma<0$ and $m<\phi<M<z$, then there is a smooth periodic solution $\phi(x-ct)$ of \eqref{DGH} with $m=\min_{z\in\mathbb{R}}\phi(z)$ and $M=\max_{z\in\mathbb{R}}\phi(z).$
	\item[$(iv)$] (\textit{Smooth with decay})Whenever $\gamma<0$ and $m<\phi<M=z$, then there exists a smooth solution $\phi(x-ct)$ of \eqref{DGH} with $m=\min_{z\in\mathbb{R}}\phi(z), M=\sup_{z\in\mathbb{R}}\phi(z)$ and $\phi \uparrow M$ exponentially as $z\to \pm\infty$.
	\end{enumerate}
\end{theorem}




\section{Case $\alpha\neq 0$:}\label{Sec4}

Before proceeding with the classification for the case $\alpha\neq 0$ we need to understand the behaviour of \eqref{eq1} as $\alpha^2(c-\phi) + \gamma\to 0$. Observe that $\alpha^2(c-\phi) + \gamma\to 0$ if and only if
$$
\phi \to \tilde{c} := \frac{\alpha^2c +\gamma}{\alpha^2}
$$
and \eqref{eq1} is rewritten as
\begin{align}\label{eq1a}
\left(\phi'\right)^2=F(\phi):=\frac{\phi^2(c-c_0-\phi) + A\phi+B}{\alpha^2(\tilde{c}-\phi)}.
\end{align}

A straightforward adaptation of Lemma 4 of \cite{Len} for \eqref{eq1a} yields the following Lemma, see also \cite{YT}.

\begin{lemma}\label{lemma1}
Let $k\in \mathbb{R}$. A function $\phi\in H^1_{loc}(\mathbb{R})$ is a travelling wave solution for $\eqref{DGH}$ if and only if the following conditions hold
\begin{enumerate}
\item[$(a)$] There are disjoint open intervals $E_i$, $i\geq 1$ and a closed set $C$ such that $\mathbb{R}\setminus C = \bigcup\limits_{i=1}^{\infty}E_i$, $\phi \in C^{\infty}(E_i)$ for every $i\geq 1$ and 
\begin{align*}
&\phi(z)\neq \tilde{c}, \quad z\in \mathbb{R}\setminus C,\\
&\phi(z) = \tilde{c},\quad z \in C.
\end{align*}

\item[$(b)$] There exists an $A\in \mathbb{R}$ such that 
	\begin{enumerate}
	\item[$(i)$] For each $i\geq 1$, there is $B_i\in \mathbb{R}$ such that 
	\begin{align}\label{eq3}
	(\phi')^2 = \frac{1}{\alpha^2}\frac{B_i + A\phi +(c-c_0)\phi^2-\phi^3}{\tilde{c} -\phi}, \quad \text{for}\,\,\, z \in E_i,
	\end{align}
	$\phi \to \tilde{c}$ at any finite endpoint of $E_i$.
	\item[$(ii)$] If the Lebesgue measure of $C$ is not zero, then $A= 3\tilde{c}^2+2(c_0-c)\tilde{c}$.
	\end{enumerate}
\item[$(c)$]$(\phi-\tilde{c})^2\in W^{2,1}_{loc}$.
\end{enumerate}
\end{lemma}

The items $(a)$ and $(b)$ of Lemma \ref{lemma1} tell that any wave solutions for this case will be smooth with the exception of points where $\phi = \tilde{c}$. It means that for weak solutions $\phi$, the function $F$ will have a smooth behaviour before and after $\phi=\tilde{c}$.


Consider \eqref{eq1a} and observe that the numerator of $F$ has either one or three real zeros. If only one zero is real, then again no bounded travelling waves will exist. Suppose $F$ has three zeros $m,M$ and $z_0$ so $P(\phi) = \frac{1}{\alpha^2}(M-\phi)(\phi-m)(\phi-z_0)$ with compatibility condition $z_0 = c-c_0-M-m.$

For bounded smooth solutions $\phi$, let $m = \inf_{z\in\mathbb{R}}\phi(z)$ and $M = \sup_{z\in\mathbb{R}}\phi(z).$
	\begin{enumerate}
	\item If $z_0 < m< \phi< M < \tilde{c}$, then $F(\phi)>0$ and there exists a smooth periodic travelling wave $\phi(z)$.

	\item If $z_0 = m < \phi<M < \tilde{c}$, there is a smooth travelling wave $\phi(z)$ with $\phi\downarrow m$ exponentially as $z\to \pm \infty$.
	\end{enumerate}

For weak solutions $\phi$, the behaviour of $m$ and $M$ can be different, as the following classification shows:
	\begin{enumerate}
	\item If $z_0 < m <\phi< M = \tilde{c}$, then
	\begin{align*}
	F(\phi)= \frac{1}{\alpha^2}(\phi-m)(\phi-z)>0,
	\end{align*}
the singularity is removed and	there is a periodic peakon wave $\phi(z)$ with $m = \min_{z\in\mathbb{R}}\phi(z)$ and $M = \max_{z\in\mathbb{R}}\phi(z).$

	\item If $z_0 = m <\phi< M = \tilde{c}$, then
	\begin{align*}
	F(\phi)= \frac{1}{\alpha^2}(\phi-m)^2>0,
	\end{align*}
	one removes the singularity and there is a peaked wave $\phi(z)$ with $m = \inf_{z\in\mathbb{R}}\phi(z), M = \max_{z\in\mathbb{R}}\phi(z),$ and $\phi \downarrow m$ exponentially as $z\to \pm\infty$.

	\item If $z_0 < m <\phi< \tilde{c} < M$, then $F(\phi)>0$ and there is a periodic cusped travelling wave $\phi(z)$ of \eqref{DGH}, with $m = \min_{z\in\mathbb{R}}\phi(z), \tilde{c} = \max_{z\in\mathbb{R}}\phi(z).$

	\item If $z_0 = m <\phi< \tilde{c} <M$, then $F(\phi)>0$ and there is a cusped travelling wave $\phi(z)$ of \eqref{DGH} with $m = \inf_{z\in\mathbb{R}}\phi(z), \tilde{c} = \max_{z\in\mathbb{R}}\phi(z),$ and $\phi \downarrow m$ exponentially as $z\to\pm\infty$.
	\end{enumerate}
	
With this we have proven items $(i)$-$(vi)$ of the following theorem on the classification of bounded waves for \eqref{DGH} with $\alpha\neq 0$, with the remaining items being analogous:	

\begin{theorem}\label{teoch}
Let $\alpha\neq 0$, $\phi\in H^1_{loc}(\mathbb{R})$ be a travelling wave solution of $\eqref{DGH}$ with speed $c$ and $\tilde{c} = \displaystyle{\frac{\alpha^2c + \gamma}{\alpha^2}}$. Then it falls into one of the following cases, with $z_0=c-c_0-M-m$:
	\begin{enumerate}
	\item[$(i)$] (\textit{Smooth periodic}) If $z_0 < m < M < \tilde{c}$, there is a smooth periodic travelling wave $\phi(x − ct)$ of \eqref{DGH}, with $m = \min_{z\in\mathbb{R}}\phi(z)$ and $M = \max_{z\in\mathbb{R}}\phi(z),$.

	\item[$(ii)$] (\textit{Smooth with decay}) If $z_0 = m < M < \tilde{c}$, there is a smooth travelling wave $\phi(x −ct)$ of \eqref{DGH} with $m = \inf_{z\in\mathbb{R}}\phi(z), M = \max_{z\in\mathbb{R}}\phi(z),$ and $\phi\downarrow m$ exponentially as $z\to \pm \infty$.

	\item[$(iii)$] (\textit{Periodic peakons}) If $z_0 < m < M = \tilde{c}$, there is a periodic peaked travelling wave $\phi(x − ct)$ of \eqref{DGH}, with $m = \min_{z\in\mathbb{R}}\phi(z)$ and $M = \max_{z\in\mathbb{R}}\phi(z).$
	
	\item[$(iv)$] (\textit{Peakons with decay}) If $z_0 = m < M = \tilde{c}$, there is a peaked travelling wave $\phi(z)$ of \eqref{DGH} with $m = \inf_{z\in\mathbb{R}}\phi(z), M = \max_{z\in\mathbb{R}}\phi(z),$ and $\phi \downarrow m$ exponentially as $z\to \pm\infty$.

	\item[$(v)$] (\textit{Periodic cuspons}) If $z_0 < m < \tilde{c} < M$, there is a periodic cusped travelling wave $\phi(z)$ of \eqref{DGH}, with $m = \min_{z\in\mathbb{R}}\phi(z), \tilde{c} = \max_{z\in\mathbb{R}}\phi(z).$

	\item[$(vi)$] (\textit{Cuspons with decay}) If $z_0 = m < \tilde{c} <M$, there is a cusped travelling wave $\phi(z)$ of \eqref{DGH} with $m = \inf_{z\in\mathbb{R}}\phi(z), \tilde{c} = \max_{z\in\mathbb{R}}\phi(z),$ and $\phi \downarrow m$ exponentially as $z\to\pm\infty$.

	\item[$(i')$] (\textit{Smooth periodic}) If $z_0>M>m>\tilde{c}$, there is a smooth periodic travelling wave $\phi(z)$ of \eqref{DGH}, with $m = \min_{z\in\mathbb{R}}\phi(z), M = \max_{z\in\mathbb{R}}\phi(z),$.

	\item[$(ii')$] (\textit{Smooth with decay}) If $z_0=M>m>\tilde{c}$, there is a smooth travelling wave $\phi(z)$ of \eqref{DGH} with $m = \min_{z\in\mathbb{R}}\phi(z), M = \sup_{z\in\mathbb{R}}\phi(z)$ and $\phi\uparrow M$ exponentially as $z\to\pm\infty$.
	
	\item[$(iii')$] (\textit{Periodic peakons}) If $z_0>M>m=\tilde{c}$, there is a periodic peaked travelling wave $\phi(x − ct)$ of \eqref{DGH}, with $m = \min_{z\in\mathbb{R}}\phi(z)$ and $M = \max_{z\in\mathbb{R}}\phi(z)$.

	\item[$(iv')$] (\textit{Peakons with decay}) If $z_0=M>m=\tilde{c}$, there is a peaked travelling wave $\phi(x−ct)$ of \eqref{DGH}, with $m = \inf_{z\in\mathbb{R}}\phi(z), M = \max_{z\in\mathbb{R}}\phi(z)$ and $\phi\downarrow m$ exponentially as $z\to\pm\infty$.
	
	\item[$(v')$] (\textit{Periodic cuspons}) If $z_0 > M > \tilde{c} > m$, there is a periodic cusped travelling wave $\phi(z)$ of \eqref{DGH} with $\tilde{c} = \min_{z\in\mathbb{R}} \phi(z)$ and $M = \max_{z\in\mathbb{R}} \phi(z)$.

	\item[$(vi')$] (\textit{Cuspons with decay}) If $z_0= M > \tilde{c} > m$, there is a cusped travelling wave $\phi(z)$ of \eqref{DGH} with $\tilde{c} = \min_{z\in \mathbb{R}}\phi(z), M=\sup_{z\in\mathbb{R}} \phi(z)$ and $\phi\uparrow M$ exponentially as $z\to \pm\infty$.
	\end{enumerate}
\end{theorem}

Taking a close look at Lemma \ref{lemma1} and comparing to Lemma 4 in \cite{Len}, we observe that the zeros of \eqref{eq3} are the same of the quadrature form of the Camassa-Holm equation. Moreover, both poles are simple. This means that Theorem \ref{teoch} for $\alpha\neq 0$ must contain the same types of solutions found by Lenells \cite{Len} for $\alpha=1$ by making $c\to \tilde{c}$ in \cite{Len}. Furthermore, it shows that, for the sake of the classification, it is enough to consider $\alpha=1$.

On the other hand, with Lemma \ref{lemma1} and Theorem \ref{teoch}, we can infer that it is possible to glue a countable number of peakon and cuspon solutions to give rise to a composite wave solution. In fact, from the compatibility condition for $\phi$, we have
\begin{align*}
A = -Mm - (M+m)(c - c_0 - M- m),
\end{align*}
which corresponds to an ellipse (see \cite{Len}). For points $(M,m,z_0)$ satisfying $(iii)-(vi)$ or $(iii')-(vi')$ and corresponding to the same $A$, we can join solutions:

\begin{corollary}
Regarding the cases obtained in Theorem \ref{teoch}, we have:
\begin{enumerate}
	\item (\textit{Composite wave solutions}) Any countable number of cuspons and peakons corresponding to the same $A$ of \eqref{eq3} can be joined at points where $\phi = \tilde{c}$ to form a composite wave $\phi$. If the Lebesgue measure of the $\phi^{-1}(\tilde{c})$ is zero, then $\phi$ is a travelling wave of \eqref{DGH}.
	\item (\textit{Stumpon solutions}) For $A=3\tilde{c} + 2(c_0-c)\tilde{c}$, the composite waves are solutions of \eqref{DGH} even if the Lebesgue measure of $\phi^{-1}(\tilde{c})$ is positive. In this case, the composite waves will be consisted of a countable number of cuspons.
\end{enumerate}
\end{corollary}

\section{Conclusion}

In this paper we studied \eqref{DGH} from the point of view of types of travelling wave solutions admitted by the equation. It is shown that the choices $\alpha=0$ and $\alpha\neq 0$ in \eqref{DGH} substantially change the classification. For the case $\alpha=0$ corresponding to the KdV equation, Theorem \ref{teokdv} states that there are only smooth (periodic and with decay) solutions. Conversely, for the case $\alpha\neq 0$ we showed in Theorem \ref{teoch} that it is possible to guarantee existence of both smooth solutions and weak solutions such as peakon and cuspon solutions. Furthermore, for points in a certain ellipse, it is possible to glue weak solutions to give rise to composite wave solutions.

\section{Acknowledgments}

The author would like to thank CAPES for her post-doctoral fellowship, and Professor Igor Leite Freire for all the fruitful suggestions and conversations regarding this paper.

\end{document}